\documentclass[runningheads]{llncs}

\usepackage{amsfonts}
\usepackage{epsfig}
\usepackage{amsmath}
\usepackage{graphicx}
\usepackage{amsmath}
\usepackage{amscd}
\usepackage{stmaryrd}
\usepackage{algorithm}
\usepackage{arydshln}
\usepackage{algorithmic}
\usepackage{youngtab}
\usepackage{young}
\usepackage{geometry}
\usepackage{bbm}\usepackage{multirow}
\usepackage{leftidx}\usepackage{mathdots}
\usepackage{bm}
\usepackage{extarrows}

\usepackage[colorlinks,
            linkcolor=blue,
            anchorcolor=blue,
            citecolor=blue,
            ]{hyperref}

\def\bt{\begin{theorem}}
\def\et{\end{theorem}}
\def\bp{\begin{proposition}}
\def\ep{\end{proposition}}
\def\bc{\begin{corollary}}
\def\ec{\end{corollary}}
\def\bo{\begin{proof}}
\def\eo{\end{proof}}
\def\bx{\begin{example}}
\def\ex{\end{example}}
\def\br{\begin{remark}}
\def\er{\end{remark}}
\def\bl{\begin{lemma}}
\def\el{\end{lemma}}

\def\bn{\begin{definition}}
\def\en{\end{definition}}
\def\ba{\begin{array}}
\def\ea{\end{array}}
\def\be{\begin{equation}}
\def\ee{\end{equation}}
\def\bd{\begin{description}}
\def\ed{\end{description}}
\def\bu{\begin{enumerate}}
\def\eu{\end{enumerate}}
\def\bi{\begin{itemize}}
\def\ei{\end{itemize}}

\newbox\bigstrutbox
\setbox\bigstrutbox=\hbox{\vrule height12.5pt depth5pt width0pt}
\def\bigstrut{\relax\ifmmode\copy\bigstrutbox\else\unhcopy\bigstrutbox\fi}

\newbox\Bigstrutbox
\setbox\Bigstrutbox=\hbox{\vrule height17.5pt depth5pt width0pt}
\def\Bigstrut{\relax\ifmmode\copy\Bigstrutbox\else\unhcopy\Bigstrutbox\fi}

\def\i{{\bf i}}

\def\0{{\bf 0}}
\def\1{{\bf 1}}
\def\2{{\bf 2}}
\def\3{{\bf 3}}
\def\4{{\bf 4}}
\def\5{{\bf 5}}
\def\6{{\bf 6}}
\def\7{{\bf 7}}
\def\8{{\bf 8}}
\def\9{{\bf 9}}

\def\ds{\displaystyle}

\begin{document}

\pagestyle{headings}

\mainmatter

\title{Quantum Arnoldi and conjugate gradient iteration algorithm}

\author{Changpeng Shao \\
cpshao@amss.ac.cn}

\authorrunning{Changpeng Shao}

\institute{Academy of Mathematics and Systems Science, Chinese Academy of Sciences, Beijing 100190, China}

\maketitle

\begin{abstract}
Arnoldi method and conjugate gradient  method are important classical iteration methods in solving linear systems and estimating eigenvalues. Their efficiency often affected by the high dimension of the space, where quantum computer can play a role in.
In this work, we establish their corresponding quantum algorithms. To achieve high efficiency, a new method about linear combination of quantum states will be proposed. The final complexity of quantum Arnoldi  iteration method is $O(m^{3+\log (m/\epsilon)}(\log n)^2 /\epsilon^4)$ and the final complexity of quantum conjugate gradient iteration method is $O(m^{1+\log m/\epsilon} (\log n)^2 \kappa/\epsilon)$, where $\epsilon$ is precision parameter, $m$ is the iteration steps, $n$ is the dimension of space and $\kappa$ is the condition number of the coefficient matrix of the linear system the conjugate gradient method works on. Compared with the classical methods, whose complexity are $O(mn^2+m^2n)$ and $O(mn^2)$ respectively, these two quantum algorithms provide us more efficient methods to solve linear systems and to compute eigenvalues and eigenvectors of general matrices.
Different from the work \cite{rebentros}, the complexity here is almost polynomial in the iteration steps. Also this work is more general than the iteration method considered in \cite{kerenidis}.
\end{abstract}


\section{Introduction}
\setcounter{equation}{0}

Quantum computer outperforms the classical computer due to its quantum property \cite{nielsen}, like parallelism, superposition, entanglement and so on. Correspondingly, quantum algorithm performs much better than the classical algorithm in many problems, like factoring \cite{shor}, searching \cite{grover}, linear system solving \cite{harrow}, to name a few. At the same time, these already discovered quantum algorithms provide us some techniques to design new quantum algorithms. For instance, from Shor's factoring algorithm, we have the quantum phase estimation method; From Grover's searching algorithm, we have amplitude amplification and quantum walk; The linear system solving algorithm given by Harrow, Hassidim and Lloyd \cite{harrow} provide us a method to study quantum machine learning. Until now, many quantum algorithms were obtained based on these techniques with great performance \cite{ambainis}, \cite{biamonte}, \cite{childs-linear-system}, \cite{childs-dam}, \cite{hallgren}, \cite{kuperberg}, \cite{magniez}, \cite{dam},  etc..

Many good quantum algorithms or techniques applies one or two ``things" belongs to quantum computer itself. And we believe that to achieve high efficiency, any quantum algorithm should follow some quantum properties.
For instance, the swap test considered in \cite{buhrman}, which can be used to estimate the inner product of two quantum states. Different from the classical inner product estimation method, swap test is a byproduct of quantum phase estimation and Grover's searching algorithm. Also the HHL algorithm to solve linear system $Ax=b$, it applies Hamiltonian simulation technique and quantum phase estimation to estimation the eigenvalues and eigenvectors of $A$ from $e^{-\i At}$. The classical method will never do this, since the exponential $e^{-\i At}$ is not easy to obtain. The singular value estimation proposed by Kerenidis and Prakash in \cite{kerenidis-prakash} was obtained by a totally different unitary matrix from the idea of quantum walk. There are also many other similar quantum algorithms that uses one or two quantum properties. So in order to find more efficient quantum algorithms from the inspiration of classical algorithms, we should make sure that the obtained quantum algorithms have followed some rules of quantum computer.

There are at least two ways to generalize classical algorithms into quantum algorithms. The first one is based on oracle, that is the unitary operator $U_f:|x,y\rangle\mapsto|x,y\oplus f(x)\rangle$. Such a generalization provides no improvement on the efficiency, however, it tells us that quantum computer can do anything the classical computer does. Another method applies the quantum algorithms or techniques in each step of classical algorithms. For instance, the iteration methods considered in \cite{kerenidis}, \cite{rebentros}. The algorithms are classical, however, quantum algorithms and techniques can make each step more efficient than apply the classical methods directly. And finally this improves the efficiency of the whole algorithm. The quantum algorithms constructed in this way also have its own drawbacks, such as the complexity of obtained quantum algorithm is exponentially depending on the iteration step, however, the complexity of the classical iteration algorithm is polynomial in iteration steps. Although iteration step is quite small in many cases, it is still hard to say that the corresponding quantum algorithm is better than the classical algorithm in its most general case.

Iterative method is an important numerical computing method that can solve many problems in mathematics efficiently.  In the past two years, people start to pay attention to generalize classical iterative methods to the corresponding quantum algorithms. In \cite{rebentros}, Rebentrost et al considered the quantum gradient descent and Newton's method for polynomial optimization problem in the high dimensional sphere. The final complexity is exponential in the iteration steps, but polynomial in the dimension of space. This is due to the No-Clone theorem, so exponential many copies of the initial states are required. In \cite{kerenidis}, Kerenidis et al studied some types of stationary linear iteration in the form $x\mapsto Ax+b$ for some fixed matrix $A$ and vector $b$. Such iteration method is very simple, but it plays important role in solving linear systems. The good point of this quantum algorithm is that it is polynomial in the iteration steps and the dimension of space. However, as we know, in most cases, the iteration method (such as in optimization, programming) are not stationary, it renews some information in each iteration. So the exponential dependence on the iteration steps seems unavoidable in quantum computer if we generalize the classical iteration method directly. Notice that one specific goal of quantum iteration method is to reduce the dependence on the space dimension, which can achieved by the quantum linear algebraic technique, since most matrix operations in quantum computer are polynomial in the space dimension.

In this paper, we consider the quantum version of Arnoldi iteration and conjugate gradient method. The classical Arnoldi iteration method \cite{arnoldi}, \cite{saad} in one of the most important iteration method of the Krylov method or projection method. It can be used to solve large sparse linear system and also can be used to estimate eigenvalues of non symmetric matrices. In the special case when the matrix is symmetric, the Arnoldi iteration reduces to Lanczos iteration. As a simplification of the Lanczos iteration to solve linear system, the conjugate gradient algorithm is one elegant variant. It is one of the best known iterative techniques for solving numerical solution of particular sparse linear systems $Ax=b$, that is $A$ is symmetric and positive-definite. Conjugate gradient algorithm can also used to estimate the eigenvalues information of $A$, such as the largest and smallest eigenvalue of $A$. This could be used to compute an estimate of the condition number of $A$. Although, there exists exponential speedup quantum algorithm \cite{abrams} to estimate the eigenvalues of symmetric matrices, it may not so efficient to estimate the largest or the smallest eigenvalue, since the quantum algorithm in \cite{abrams} treat all the eigenvalues equally.

As we know, Krylov method aims at approximating $A^{-1}$ by a polynomial of $A$. When we obtain the information about the coefficients of this polynomial, then it reduces to the problem of calculating the power of matrix, which can actually be calculated efficiently in quantum computer. Just like the stationary iteration method, we also need to compute the linear combination of these matrix powers. The linear combination methods of quantum states (or unitaries) has been considered such as in \cite{berry1}, \cite{berry2}, \cite{childs} to study Hamiltonian simulations, considered in \cite{childs-linear-system} to study solving linear systems and also considered in \cite{clader} to study the preparation of quantum states. Their methods about linear combination methods of quantum states are similar to classical methods, that is try to get the linear combination of quantum states, then normalize it be measuring. The complexity depends on the linear coefficients and the 2-norm of the linear combined vector the desired quantum state proportional to. This idea works well in some cases, and these methods can also play certain roles in our study about quantum Arnoldi iteration and conjugate gradient method. In this work, however, we will propose another method to achieve linear combination of quantum states, which is more suitable in solving our problem, under the assumption that $U^r$ can be efficiently implemented for any $0\leq r\leq 1$ if $U$ does. The basic idea is simple. Since quantum states are normalized, and so they all lie in the unit sphere. We can actually perform some rotations on the given quantum states to get the desired quantum states. The complexity of such a method is independent of the linear coefficients and the norm of the desired quantum state.

As shown in section \ref{Quantum Arnoldi iteration}, the complexity of the directly generalized quantum algorithm of classical Arnoldi iteration method is exponential on the iteration steps. But the good point of this quantum algorithm is that it is independent of the condition number of the given matrix. So to solve the linear system $Ax=b$, if the iteration step $m$ is a small constant, which happens in many cases, then the complexity is $O((\log n)^{2m-2}/\delta^{m-1}\epsilon^{m-1})$, where $\delta,\epsilon$ are some fixed parameters and $n$ is the dimension of matrix. On the other hand, an improved quantum version still exists due to the property of the Krylov method. The complexity can be improved into $O((m+6)!(\log n)^2/\delta\epsilon^2)$ based on the old method to achieve linear combination of quantum states, and can further improved into $O(m^{3+\log (m/\epsilon)}(\log n)^2 /\epsilon^4)$ based on our new method to achieve linear combination of quantum states.
As for the conjugate gradient method studied in section \ref{Quantum conjugate gradient algorithm}, the complexity is $O(m^2 (\log n)^2\kappa/\delta^{3m}\epsilon)$ based on the old method and is $O(m^{1+\log m/\epsilon} (\log n)^2 \kappa/\epsilon)$ based on our new method  to achieve linear combination of quantum states.

The structure of this paper is as follows. In section \ref{Preliminary techniques}, we introduce some preliminary knowledge required in this work and the new method to achieve linear combination of quantum states. Some new applications like triangle finding problem, power iteration method will be discussed as applications of HHL algorithm.
Section \ref{Quantum stationary iteration} is denoted to study the stationary iteration method.
In section \ref{Quantum Arnoldi iteration} and section \ref{Quantum conjugate gradient algorithm}, we aim at study the quantum version of Arnoldi and conjugate gradient method.
Finally, section \ref{conclusion} is a conclusion. In this work, the norm $\|\cdot\|$ always refers to 2-norm of vectors and $\i$ refers to the imaginary unit $\sqrt{-1}$.

\section{Preliminary techniques}
\label{Preliminary techniques}
\setcounter{equation}{0}

In this section, we provide some preliminary techniques that will be used in this whole work. The first one is swap test, which can be used to evaluate the inner product of quantum states. This is quite useful in matrix-vector operations. The second one is about the quantum linear algebraic technique arises from HHL algorithm. It can achieve exponential speedup to perform simple matrix vector operation, like multiplication and inversion. Three new applications of this technique will be discussed then. The last technique will be used in this work is the linear combination of quantum states, which has been used in Hamiltonian simulation and linear systems solving. A new method with much better performance will be proposed.

\subsection{Swap test}

The following lemma is a direct corollary of quantum phase estimation and Grover iteration.

\bl \label{lem:inner product}
Let $|\phi\rangle=\sin\theta|0\rangle|u\rangle+\cos\theta|1\rangle|v\rangle$ be a unknown quantum state that can be prepared in time $O(T_{\emph{in}})$, where $|u\rangle,|v\rangle$ are normalized quantum states. Then there is a quantum algorithm that can compute $\sin\theta,\cos\theta$ in time $O(T_{\emph{in}}/\epsilon\delta)$ with accuracy $\epsilon$ and with success probability at least $1-\delta$.
\el

\bo
Let $Y$ be the 2-dimensional unitary transformation that maps $|0\rangle$ to $-|0\rangle$ and $|1\rangle$ to $|1\rangle$. Denote $G=(2|\phi\rangle\langle\phi|-I)(Y\otimes I)$ which is the rotation matrix used in Grover's searching algorithm. Then
\[
G=\left(
   \begin{array}{rr}
     \cos2\theta  &~~ \sin2\theta \\
     -\sin2\theta &~~ \cos2\theta \\
   \end{array}
 \right)
 \]
in the basis $\{|0\rangle|u\rangle,|1\rangle|v\rangle\}$.
The eigenvalues of $G$ are
\[
e^{\i2\theta}=\cos2\theta+\i \sin2\theta,~~e^{-\i2\theta}=\cos2\theta-\i \sin2\theta
\]
and the corresponding eigenvectors are
\[
|w_1\rangle=\frac{1}{\sqrt{2}}\Big(|0\rangle|u\rangle+\i|1\rangle|v\rangle\Big),~~|w_2\rangle=\frac{1}{\sqrt{2}}\Big(|0\rangle|u\rangle-\i|1\rangle|v\rangle\Big)
\]
respectively. Note that
\[
|\phi\rangle = \ds\frac{\sin\theta}{\sqrt{2}}\Big(|w_1\rangle+|w_2\rangle\Big)-\frac{\i \cos\theta}{\sqrt{2}}\Big(|w_1\rangle-|w_2\rangle\Big)
= \ds-\frac{\i}{\sqrt{2}}\Big(e^{\i\theta}|w_1\rangle-e^{-\i\theta}|w_2\rangle\Big).
\]
So performing quantum phase estimation algorithm on $G$ with initial state $|0\rangle^n|\phi\rangle$, for some $n=O(\log1/\delta\epsilon)$, can help us find an approximation $\tilde{\theta}$ of $\theta$ with failure probability $\delta$, such that $|\tilde{\theta}-\theta|\leq \epsilon$. Then it is easy to check $|\sin\theta-\sin\tilde{\theta}|\leq \epsilon$.
\eo

Quantum counting \cite{brassard} also share the same essence. Actually, the quantum phase estimation algorithm on $G$ in the above proof returns a state in the form
\be\label{final-form}
-\frac{\i}{\sqrt{2}}\Big(e^{\i\theta}|y\rangle|w_1\rangle-e^{-\i\theta}|-y\rangle|w_2\rangle\Big),
\ee
where $y\in \mathbb{Z}_{2^n}$ satisfies $|\theta-y\pi/2^n|\leq \epsilon$. Let $f(y)=g(\theta)$ be some functions such that $f(y)=f(-y)$, then from \eqref{final-form}, we can get
\be\label{final-form-1}
|g(\theta)\rangle|\phi\rangle.
\ee
Generally, the failure probability $\delta$ can be ignored. A directly corollary of lemma \ref{lem:inner product} is the following result, and it is usually called the swap test \cite{buhrman}.

\bc\label{cor:inner product}
Let $|x\rangle,|y\rangle$ be two real quantum states, which can be prepared in time $O(T_{\emph{in}})$,
then $\langle x|y\rangle$ can be estimated with accuracy $\epsilon$ in time $O(T_{\emph{in}}/\epsilon)$.
\ec

\bo Consider the state $|\phi\rangle=\frac{1}{\sqrt{2}}(|+\rangle|x\rangle+|-\rangle|y\rangle)$.
Then the probability of the first qubit is $|0\rangle$ (resp. $|1\rangle$) is $(1+\langle x|y\rangle)/2$ (resp. $(1-\langle x|y\rangle)/2$).
By lemma \ref{lem:inner product}, these two values can be evaluated in time $O(T_{\textmd{in}}/\epsilon)$ with accuracy $\epsilon$.
Then so is $\langle x|y\rangle$.
\eo

Moreover, from \eqref{final-form-1}, we actually can obtain the following quantum state
\be\label{final-form-2}
\frac{1}{\sqrt{2}}\Big(|0\rangle|x\rangle+|1\rangle|y\rangle\Big)\Big|g(\langle x|y\rangle)\Big\rangle,
\ee
for any function $g$, because $\cos\theta$ is an even function. Therefore, we have

\bp
Let $|x\rangle,|y\rangle$ be two real quantum states, except a global phase, which can be prepared in time $O(T_{\emph{in}})$. Let $f$ be any function. Then there is a quantum algorithm within time $O(T_{\emph{in}}/\epsilon)$ to achieve
\be\label{final-form-3}
\frac{1}{\sqrt{2}}(|0\rangle|x\rangle+|1\rangle|y\rangle)\mapsto
\frac{1}{\sqrt{2}}(|0\rangle|x\rangle+|1\rangle|y\rangle)|f(s)\rangle,
\ee
where $|\langle x|y\rangle-s|\leq \epsilon$.
\ep

This result tells us that, we can put the inner product of $|x\rangle$ and $|y\rangle$ into quantum state as quantum information.
This is important in the case when we need to parallelly deal with the inner product of quantum states in quantum computing, such as matrix multiplication.
If $|x\rangle,|y\rangle$ are complex quantum states, then the probability of $|0\rangle$ (resp. $|1\rangle$) is $(1+\textmd{Re}\langle x|y\rangle)/2$ (resp. $(1-\textmd{Re}\langle x|y\rangle)/2$). So, we can only get the value of $\textmd{Re}\langle x|y\rangle$.
The image part of $\langle x|y\rangle$ can be computed by considering the inner product of $|x\rangle$ with $\i|y\rangle$.

\subsection{Quantum matrix-vector multiplication}

Let $A$ be a $n\times n$ Hermitian matrix and $|y\rangle$ a given quantum state. Assume that $f$ is a map in one variable. Then we can obtain the quantum state proportional to $f(A)|y\rangle$ efficiently by a similar procedure as HHL algorithm \cite{harrow}. Due to the requirements appeared in HHL algorithm, here we make the following assumptions.

\begin{description}
  \item[Assumption 1] The Hamiltonian simulation $e^{-\i At}$ can be implemented efficiently in time $\widetilde{O}(t)$.
\end{description}

The Hamiltonian simulation can be solved efficiently in some cases \cite{childs}. Assumption 1 is not so necessary when considering HHL algorithm to solve linear systems, since in \cite{kerenidis-prakash}, Kerenidis et al proposed a new version of singular value estimation (SVE) method, which is independent of Hamiltonian simulation. However, in this work, we choose to follow the way of HHL algorithm, since the SVE studied in \cite{kerenidis-prakash} also contain other assumptions. Suppose the singular value decomposition of $A=\sum_{j=1}^n\lambda_j|u_j\rangle\langle u_j|$ and $|y\rangle=\sum_{j=1}^n\beta_j|u_j\rangle$, then $f(A)|y\rangle=\sum_{j=1}^nf(\lambda_j)\beta_j|u_j\rangle$.

\begin{description}
  \item[Assumption 2] The singular values of $A$ lie between $1/\kappa$ and 1, where $\kappa$ is the condition number of $A$.
\end{description}

This assumption is not so necessary too, since when we have some better upper bound information about the singular values of $A$, on one hand, we can perform a scaling in advance; and on the other hand, such a scaling can be performed in the quantum phase estimation algorithm.

\begin{description}
  \item[Assumption 3] If $f(\lambda_j)=0$ then $\beta_j=0$.
\end{description}

This assumption is quite important, which relates to the success probability. It is equivalent to assume that $|y\rangle$ lies in the non-zero (well-conditioned) components of $f(A)$. When $A$ is invertible and $f$ is a simple function like $f(x)=x^l$, then assumption 3 is unnecessary.

The following procedures are similar to HHL algorithm. So we just briefly review it. In the quantum phase estimation algorithm, we choose the initial state as $|0\rangle|y\rangle$, then it yields an approximates state
$
\sum_{j=1}^n\beta_j|\tilde{\lambda}_j\rangle|u_j\rangle
$
in time $\widetilde{O}(1/\epsilon)$, where $\epsilon$ is the accuracy to estimate the singular values, that is $|\lambda_j-\tilde{\lambda}_j|\leq \epsilon$. Add an ancilla register to the obtained state and apply a control rotation based on the register stores singular value, undo the quantum phase estimation algorithm, we have
\be\label{hhl:eq2}
\sum_{j=1}^n\beta_j|u_j\rangle\Big(f(\tilde{\lambda}_j)C|0\rangle+\sqrt{1-f(\tilde{\lambda}_j)^2C^2}|1\rangle\Big),
\ee
where $C$ is a constant such that $|f(\tilde{\lambda}_j)C|\leq 1$, for instance we can choose $C=1/\max_j|f(\tilde{\lambda}_j)|$. Because of assumption 3, the probability to obtain the quantum state of $f(A)|y\rangle$ is
\be\label{hhl:eq3}
\sum_{j=1}^n\Big|\beta_j f(\tilde{\lambda}_j)C\Big|^2\geq \frac{\min_{j:f(\tilde{\lambda}_j)\neq 0}|f(\tilde{\lambda}_j)|^2}{\max_j|f(\tilde{\lambda}_j)|^2},
\ee
Finally, the complexity to get the quantum state of $f(A)|y\rangle$ should be multiplied by
\be\label{hhl:eq4}
\max_j|f(\tilde{\lambda}_j)|/\min_{j:f(\tilde{\lambda}_j)\neq 0}|f(\tilde{\lambda}_j)|
\ee
due to amplitude amplification technique.
Note that in the analysis above, assumption 3 is very important. If assumption 3 fails, then the success probability estimation given in formula \eqref{hhl:eq3} will be affected.

In the following, we consider a special case of $f$ that will be used in the paper. Note that, if we only interested in the vector $f(A)|y\rangle$ before normalization, then sometimes \eqref{hhl:eq2} is enough for our analysis, which is not too expensive to get it.

Assume that $f(x)=x^l$ for some $l\in\mathbb{N}^*$. At this time, we can just choose $C=1$. Because of assumption 2, we have $|\lambda_j^l-\tilde{\lambda}_j^l|\leq l \epsilon$. Denote
\be\label{hhl:eq5}
|\psi\rangle=\frac{1}{\sqrt{Z}}\sum_{j=1}^n\lambda_j^l\beta_j|u_j\rangle,~~
|\tilde{\psi}\rangle=\frac{1}{\sqrt{\widetilde{Z}}}\sum_{j=1}^n\tilde{\lambda}_j^l\beta_j|u_j\rangle,
\ee
where
$
Z=\sum_{j=1}^n|\lambda_j^l\beta_j|^2$ and $\widetilde{Z}=\sum_{j=1}^n|\tilde{\lambda}_j^l\beta_j|^2.
$
This means $|\tilde{\psi}\rangle$ is the post measurement state by getting $|0\rangle$ in \eqref{hhl:eq2}, and $|\psi\rangle$ is the normalized target state of $A^l|y\rangle$.
Then
\be\label{hhl:eq7}
\Big|Z-\widetilde{Z}\Big| \leq  \sum_{j=1}^n\Big|\beta_j\Big|^2\Big|\lambda_j^{2l}-\tilde{\lambda}_j^{2l}\Big|\leq 2l\epsilon.
\ee
And
\be\label{hhl:eq8}
\Big|\sqrt{Z}-\sqrt{\widetilde{Z}}\Big| =  \frac{\Big|Z-\widetilde{Z}\Big|}{\Big|\sqrt{Z}+\sqrt{\widetilde{Z}}\Big|}\leq \frac{l\epsilon}{\sqrt{\widetilde{Z}}}.
\ee
In the estimation of the above inequality, we simply assume that $\widetilde{Z}\leq Z$, this does not affect the analysis below if the contrary holds.
Note that $1\geq Z,\widetilde{Z}\geq \kappa^{-2l}$, so
\be\ba{lll} \vspace{.2cm}\label{hhl:eq9}
 \ds \Big\||\psi\rangle-|\tilde{\psi}\rangle\Big\|^2
&=& \ds\frac{1}{Z\widetilde{Z}}\sum_{j=1}^n\Big|\beta_j\Big|^2\Big|\sqrt{\widetilde{Z}}\lambda_j^{l}-\sqrt{Z}\tilde{\lambda}_j^{l}\Big|^2 \\ \vspace{.2cm}
&\leq& \ds  \frac{1}{Z\widetilde{Z}}\sum_{j=1}^n\Big|\beta_j\Big|^2\Big|\sqrt{\widetilde{Z}}\Big|\lambda_j^{l}-\tilde{\lambda}_j^{l}\Big|
       +\tilde{\lambda}_j^{l}\Big|\sqrt{\widetilde{Z}}-\sqrt{Z}\Big|\Big|^2 \\\vspace{.2cm}
&\leq& \ds  \frac{l^2\epsilon^2}{Z\widetilde{Z}}\Big|\sqrt{\widetilde{Z}}+\frac{1}{\sqrt{\widetilde{Z}}}\Big|^2
= l^2\epsilon^2\Big(\frac{1}{Z}+\frac{2}{Z\widetilde{Z}}+\frac{1}{Z\widetilde{Z}^2}\Big) \\
&\leq& \ds  l^2\epsilon^2(\kappa^{2l}+2\kappa^{4l}+\kappa^{6l})=O(l^2\kappa^{6l}\epsilon^2).
\ea\ee
To make sure that the measured state $|\tilde{\psi}\rangle$ is a good approximate of $|\psi\rangle$, the error $\epsilon$ should be chosen as $\epsilon/l\kappa^{3l}$. Together with the amplitude amplification complexity \eqref{hhl:eq3}, the total complexity to get a good approximate of $|\psi\rangle$ is $\widetilde{O}(l\kappa^{4l}/\epsilon)$.

Note that if $|x\rangle$ is another quantum state, and we want to estimate $\langle x|A^l|y\rangle$ to some accuracy $\epsilon$, we actually do not need to perform a measurement in \eqref{hhl:eq2}. Instead, we can apply swap test \cite{buhrman} (also see corollary \ref{cor:inner product}) to estimate the inner product of $|x,0\rangle$ with the state \eqref{hhl:eq2}. Note that before normalization, from formula \eqref{hhl:eq5}, we have
\be \label{error}
\Big\|A^l|y\rangle-\sqrt{\widetilde{Z}}|\tilde{\psi}\rangle\Big\| \leq l\epsilon.
\ee
So the complexity to estimate $\langle x|A^l|y\rangle$ equals $\widetilde{O}(l/\epsilon^2)$, which is independent of the condition number.

When $A$ is not Hermitian, then all the results discussed above can be obtained similarly by considering the extended matrix
$\widetilde{A}=\left(
   \begin{array}{cc}
     0 & A \\
     A^\dagger & 0 \\
   \end{array}
 \right)$. Concluding above analysis, we have

\bp \label{thm1}
Let $A$ be a matrix such that $\widetilde{A}$ satisfies assumption 1-3. Let $|x\rangle$ and $|y\rangle$ be two given quantum states. Assume that $l\in\mathbb{N}^*$. Then

(1). The quantum state of $A^l|y\rangle$ can be obtained in time $\widetilde{O}(l\kappa^{4l}/\epsilon)$ to accuracy $\epsilon$.

(2). The inner product $\langle x|A^l|y\rangle$ can be estimated in time $\widetilde{O}(l/\epsilon^2)$ to accuracy $\epsilon$.

Where $\kappa$ is the condition number of $A$ and $\epsilon$ is the accuracy. Moreover, if the quantum state $|x\rangle$ and $|y\rangle$ are prepared in time $O(T_{\emph{in}})$, then the complexity should be multiplied by $O(T_{\emph{in}})$.
\ep

There may exist some other better ways to do the complexity analysis above, however, the influence of condition number cannot removed. Note that in the classical case, the complexity to compute $A^l|y\rangle$ is polynomial in $l$, but exponential in quantum computer. As we can see in the above analysis, the influence of condition number comes from the estimation of the norm of $\|A^l|y\rangle\|$. We choose using the low bound $\kappa^{-l}$. It may happen that the norm $\|A^l|y\rangle\|$ ia not small in that size in some specific problems. Such a worst case estimation makes the complexity of quantum algorithm looks ``very bad".
In the following, we show three applications of the above result: \vspace{.2cm}

\textbf{Application 1: triangle finding}

First, we provide a simple application of the above result in the triangle finding problem in graph theory. This problem has been studied a lot based on quantum walk and many improvements were obtained in the past. It can be viewed as the simplest case of the clique problem. In the following, we will show that it can be solved by quantum linear algebra based algorithm with optimal complexity in sparse case.

Given a undirected weightless graph $G=(V,E)$ with $n=\#(V)$, the triangle finding problem aims at deciding and finding three vertices $i, j, k\in V$ such that $(i,j), (j,k), (k,i)\in E$. The quantum algorithm to this problem has been considered in a lot of works.
The current best quantum algorithm to this problem is obtained by Le Gall et al. \cite{legall}, which has complexity $\widetilde{O}(n^{1.25})$. The low bound of quantum algorithm to this problem is $\Omega(n)$. Moreover, when the graph is sparse in size $\widetilde{O}(n)$, then the result of Le Gall et al. \cite{legall} can achieve the low bound $O(n)$.

This problem can be reduced to consider the diagonal entries of the cubic power of the adjacent matrix of the graph. The adjacent matrix $A=(a_{ij})_{n\times n}$ of $G$ is defined as $a_{ij}=1$ if $(i,j)\in E$ and $a_{ij}=0$ if $(i,j)\not\in E$. Then deciding whether or not there is a triangle in the graph is equivalent to deciding whether or not there is a nonzero diagonal entry of $B=A^3=(b_{ij})_{n\times n}$. If $b_{ii}$ is nonzero, then there exist $j,k\in V$ such that $i,j,k$ form a triangle. By Grover's searching algorithm, $j$ and $k$ can be found in time $O(n)$. By the result in proposition \ref{thm1}, each diagonal entry of $B$ can be evaluated in time $\widetilde{O}(1/\epsilon^2)$. So in quantum computer, we can first find a nonzero diagonal entry of $B$, which will cost at most $\widetilde{O}(n/\epsilon^2)$. When obtaining such a nonzero diagonal entry, we stop and apply Grover's searching algorithm to find the triangle in time $O(n)$. Therefore, the total complexity of the triangle finding problem is $\widetilde{O}(n/\epsilon^2)$. In this algorithm, we only need to assume that $A$ is sparse or the Hamilton simulation of $A$ can be efficiently implemented. Therefore, this quantum algorithm also achieves the low bound when the graph is sparse.

Note that this quantum algorithm also works to find $l$ polygon, not just triangle. Deciding the existence of this polygon takes time $\widetilde{O}(l^2n/\epsilon^2)$. It is also not hard to find such a polygon. Since we spend $\widetilde{O}(ln/\epsilon^2)$ to find a vertex $i$ of the polygon. Then we can consider the subgraph generated by neighbors of $i$. In this subgraph, we can find another vertex $j\neq i$ of the polygon. This also takes time $\widetilde{O}(ln/\epsilon^2)$. Continue this procedure, after $l$ steps, we can find the polygon. The total complexity is $O(l^2n/\epsilon^2)$. Also since the $(i,j)$-th entry of $A^l$ gives the number of walks of length $l$ from vertex $i$ to vertex $j$. This can be decided in time $\widetilde{O}(l/\epsilon^2)$ in quantum computer when the graph is sparse.

\bc
Let $G$ be a graph with $n$ vertex such that the Hamiltonian simulation of its adjacent matrix can be efficiently implemented, then there exist quantum algorithms

(1). to find one $l$ polygon in time $O(l^2n(\log n)^2/\epsilon^2)$,

(2). to find the number of walks of length $l$ among two given vertexes in time $O(l(\log n)^2/\epsilon^2)$.
\ec

\vspace{.2cm}

\textbf{Application 2: classical matrix multiplication}

Next, we give a brief note about the quantum linear algebra method to achieve matrix multiplication. Given two $n\times n$ matrices $A,B$. Suppose that $A, B$ satisfy assumptions 1-2, then the $(i,j)$-th entry of $AB$ equals $\langle i|AB|j\rangle$, that is the inner product of $A|i\rangle$ and $B|j\rangle$. This value can be obtained directly by applying swap test to two types of quantum state in the form \eqref{hhl:eq2}. The complexity will be $\widetilde{O}(1/\epsilon^2)$. And so the total complexity to achieve matrix multiplication in this way is $\widetilde{O}(n^2/\epsilon^2)$.

\bc
The multiplication of two sparse matrices that satisfy assumptions 1 and 2 can be achieved in time $\widetilde{O}(n^2/\epsilon^2)$ to accuracy $\epsilon$ in quantum computer.
\ec

\vspace{.2cm}

\textbf{Application 3: power iteration method}

In numerical computing, power iteration \cite{burden} is an algorithm that can approximate the greatest (in absolute value) eigenvalue (also called dominant eigenvalue) of a diagonalizable sparse matrix $A$. At the same time, it returns the corresponding eigenvector (called dominant eigenvector). The power iteration algorithm starts with an initial unit vector $|b_0\rangle$, which may be an approximate of the dominant eigenvector. The algorithm is described by the following iterative relation
\be \label{power-iteration:eq1}
|b_{k+1}\rangle=\frac{A|b_k\rangle}{\|A|b_k\rangle\|}=\frac{A^{k+1}|b_0\rangle}{\|A^{k+1}|b_0\rangle\|}.
\ee
This algorithm works under the following assumption.

\begin{description}
  \item[Assumptions 4] The matrix $A$ has only one dominant eigenvalue and the initial state $|b_0\rangle$ has a nonzero component in the dominant eigenvector.
\end{description}

Under the above assumption, it can be proved that the sequence $\{|b_k\rangle:k=0,1,\ldots\}$ converges to the dominant eigenvector. Assume that $A$ is a $n\times n$ sparse matrix with eigenvalues $\lambda_1,\lambda_2,\ldots,\lambda_n$, where $1\geq |\lambda_1|\geq |\lambda_2|\geq\ldots\geq |\lambda_n|$. The iteration step of power iteration algorithm is $\eta\leq (\log\epsilon)/\log(|\lambda_2/\lambda_1|)$, where $\epsilon$ is the estimating error of dominant eigenvector. So the classical algorithm needs $O(\eta n^2)$ arithmetic operations.
If considering $A$ directly in the power iteration algorithm, then the quantum algorithm takes $\widetilde{O}(\eta\kappa^{4\eta}/\epsilon)$. This is not good when the condition number of $A$ is large. In the following, we provide a method to decrease the dependence on condition number.

For simplicity, we assume that $A$ is Hermitian, and the dominant eigenvalue is positive. Denote $B=A+\mu I$ for some $\mu>0$ decided later. Then it has the same dominant eigenvector as $A$, since we have assumed that $\lambda_1>0$, if it is negative, we can choose $\mu<0$. The condition number of $B$ is
\be
\tilde{\kappa}=\left|\frac{\lambda_1+\mu}{\lambda_i+\mu}\right|\leq \frac{\mu+1}{\mu-1}=1+\frac{2}{\mu-1}.
\ee

The iteration step of power iteration algorithm on $B$ is $\tilde{\eta}=(\log\epsilon)/\log(|(\lambda_j+\mu)/(\lambda_1+\mu)|)$ for some $j$.
For simplicity, in the analysis of the relationship between $\tilde{\eta}$ and $\eta$, we assume that $\lambda_1=1$. And we consider the relationship between $\eta$ and $\tilde{\eta}$ in the worst case. Denote
$|\lambda_2|=1-\delta_2$, then the worst case is $\delta_2\ll 1$ and so
\be
\eta=\frac{\log\epsilon}{\log(1-\delta_2)}\approx \frac{-\log\epsilon}{\delta_2}.
\ee
If $\lambda_j=1-\delta_j>0$, then $\delta_2\leq\delta_j$ and in the the worst case $\delta_j\ll 1$, we have
\be
\tilde{\eta}=\frac{\log\epsilon}{\log(1-\delta_j/(\mu+1))}\approx\frac{-(\mu+1)\log\epsilon}{\delta_j}\leq (\mu+1)\eta.
\ee
In this case, the complexity is
\be\label{complexity:case1}
\widetilde{O}((\mu+1)\eta(1+2/(\mu-1))^{(\mu+1)\eta}/\epsilon).
\ee
If $\lambda_j=-1+\delta_j<0$, then in the worst case $\delta_2\leq\delta_j\ll 1$,
\be
\tilde{\eta} =\ds \frac{\log\epsilon}{\log((\mu-1+\delta_j)/(\mu+1))} \approx \ds\frac{-\log\epsilon}{\log(1+1/\mu)}
\approx \ds \frac{\delta_2}{\log(1+1/\mu)}\eta.
\ee
In this case, the complexity is
\be\label{complexity:case2}
\widetilde{O}\Big(\frac{\delta_2\eta}{\epsilon\log(1+1/\mu)}\Big(1+\frac{2}{\mu-1}\Big)^{\frac{\delta_2}{\log(1+1/\mu)}\eta}\Big).
\ee
In each case, the iteration step is changed into a scalar of the original iteration step. If we choose $\mu$ as a small integer, then the complexity can be simply rewritten as
$\widetilde{O}(\eta c_1^\eta/\epsilon)$ for some small constant $c_1$. From \eqref{complexity:case1} and \eqref{complexity:case2}, we see that $c_1\approx e^2\approx 7.389$.
Compared with the original quantum power iteration algorithm with $A$, the condition number now is changed into a constant.

As for the dominant eigenvalue, the complexity can be better. Since $|b_\eta\rangle$ is a good approximate of the dominant eigenvector, then we have $A|b_\eta\rangle\approx \lambda_1 |b_\eta\rangle$, which means $\lambda_1\approx\langle 0|A|b_\eta\rangle/\langle 0|b_\eta\rangle=\langle 0|A^{\eta+1}|b_0\rangle/\langle 0|A^{\eta}|b_0\rangle$. From proposition \ref{thm1}, this can be evaluated in time $\widetilde{O}(\eta/\epsilon^2)$.

\bc
The power iteration method, if convergent, works in time $\widetilde{O}(\eta c_1^\eta/\epsilon)$ to find the dominant eigenvector and in time $\widetilde{O}(\eta/\epsilon^2)$ to find the dominant eigenvalue.
\ec

For example, the Laplacian matrix $L$ of a given simple graph. We have $\lambda_1\geq ns/(n-1)$ and $\lambda_2\leq 2s$, where $s$ is the sparsity of $L$. So $\lambda_1/\lambda_2\geq 2n/(n-1)$. Assume that $n$ is large, then
\[
\eta = \ds\frac{-\log \epsilon}{\log (\lambda_1/\lambda_2)} \leq \frac{-\log \epsilon}{\log 2n/(n-1)} \approx \frac{-\log \epsilon}{1+(n-1)^{-1}}
     \leq \log(1/\epsilon).
\]
So the complexity of the quantum power iteration algorithm to find the dominant eigenvector of $L$ is
$
\widetilde{O}(\eta c_1^\eta/\epsilon)=\widetilde{O}((1/\epsilon)^3\log(1/\epsilon)).
$

\subsection{Linear combination of quantum states}

The following result about linear combination of quantum states comes from one step of HHL algorithm.

\bl \label{lem:lin-com}
Given $l$ complex numbers $\alpha_j=r_je^{\i\theta_j}$ and $l$ quantum states $|v_j\rangle$, where $j=0,1,\ldots,$ $l-1$, then we can obtain the quantum state proportional to $v_l=\sum_{j=0}^{l-1}\alpha_j|v_j\rangle$ in time
\be \label{complexity:lin-com}
O((T_{\emph{in}}+\log l)\max_{0\leq j\leq l-1}  |\alpha_j|l/\|v_l\|),
\ee
where $T_{\emph{in}}$ is the maximal complexity to prepare $|v_j\rangle$ for $j=0,1,\ldots,l-1$.
\el

\bo
The procedure is quite simple as follows
\be\ba{lll}\vspace{.2cm} \label{lin-com:method-2}
\ds\frac{1}{\sqrt{l}}\sum_{j=0}^{l-1}|j\rangle|0,0\rangle
 &\mapsto& \ds\frac{1}{\sqrt{l}}\sum_{j=0}^{l-1}|j\rangle|v_j\rangle|0\rangle \\\vspace{.2cm}
 &\mapsto& \ds\frac{1}{\sqrt{l}}\sum_{j=0}^{l-1}|j\rangle|v_j\rangle\Big(t\alpha_j|0\rangle+\sqrt{1-t^2|\alpha_j|^2}|1\rangle\Big) \\
 &\mapsto& \ds\frac{1}{l}|0\rangle|v_j\rangle\Big(t\alpha_j|0\rangle+\sqrt{1-t^2|\alpha_j|^2}|1\rangle\Big)+|0\rangle^\bot,
\ea\ee
where $t=1/\max_j|\alpha_j|$. The success probability is $\|v_l\|^2/\max_j|\alpha_j|^2l^2$. The complexity to get the desired state is
$O((T_{\textmd{in}}+\log l)\max_j|\alpha_j|l/\|v_l\|)$.
\eo

Another method to solve the problem in lemma \ref{lem:lin-com} is as follows \cite{childs}:
Denote $s=\sum_{j=0}^{l-1}|\alpha_j|$. Define unitary transformation $U$ as $U|0\rangle=\frac{1}{\sqrt{s}}\sum_{j=0}^{l-1}\sqrt{r_j}|j\rangle$.
%Then the first row of $U^\dagger$ equals $(\sqrt{r_0/s},\cdots,\sqrt{r_{l-1}/s})$, which means $U^\dagger|j\rangle=\sqrt{r_j/s}|0\rangle+\cdots$.
Consider the following procedure:
\be\ba{lcl}\vspace{.2cm} \label{lin-com:method-1}
|0\rangle|0\rangle &\xrightarrow[]{U\otimes I}& \ds\frac{1}{\sqrt{s}} \sum_{j=0}^{l-1} \sqrt{r_j}|j\rangle|0\rangle \\\vspace{.2cm}
&\rightarrow& \ds\frac{1}{\sqrt{s}} \sum_{j=0}^{l-1} \sqrt{r_j}e^{\i\theta_j}|j\rangle|v_j\rangle \\
&\xrightarrow[]{U^\dagger\otimes I}& \ds\frac{1}{s} |0\rangle \sum_{j=0}^{l-1} \alpha_j|v_j\rangle + |0\rangle^\bot
\ea\ee
The probability of the first register is $|0\rangle$ equals $\|v_l\|^2/s^2$, and so the complexity to obtain the desired quantum state is $O((T_{\textmd{in}}+C_U)s/\|v_l\|)$, where $C_U$ is the complexity to implement $U$ in quantum computer.

In the following, we provider another quantum algorithm to achieve the linear combination of quantum states. Note that the quantum state obtained by the above methods are exact, which means no error between the obtained quantum state and the desired quantum state. If we allow some error among them, then we can actually make the quantum algorithm independent of the influence of the norm $\|v_l\|$. This is quite important, since just like the analysis in HHL algorithm to solve linear system, the existence of $\|v_l\|$ increase the dependence on the condition number to quadratic.
Before the introduction of our new method, we consider the following problem first: Let $U$ be an efficiently implemented unitary operator, then does $U^t$ for any $0\leq t\leq 1$ still efficiently implemented?

Assume that the eigenvalue decomposition of $U$ as $U=\sum e^{\i\theta_j}|u_j\rangle\langle u_j|$, and let $|v\rangle=\sum \beta_j|u_j\rangle$ be any given quantum state, then
$U^t|v\rangle=\sum e^{\i\theta_j t}\beta_j|u_j\rangle$. By quantum phase estimation, this state can be obtained in time $O(C_U/\epsilon)$, where $C_U$ is the complexity to implement $U$. Therefore, we have

\bl \label{lem-lcs}
Let $U$ be an unitary operator that can be implemented in time $O(C_U)$, then $U^t$ for any $0\leq t\leq 1$ can be implemented in time $O(C_U/\epsilon)$ to accuracy $\epsilon$.
\el

Note that there may exist better ways to implement $U^t$, the above method is the most obvious one. Actually, it can be shown that the above result is optimal.
Consider the Grover searching problem $f:\mathbb{Z}_{2^n}\rightarrow \mathbb{Z}_2$ with one target $x_0$, and $f(x)=0$ if and only if $x=x_0$. Now we denote
\[\ba{lll} \vspace{.2cm}
|a\rangle &= \ds \frac{1}{\sqrt{2^n}} \sum_{x=0}^{2^n-1}|x\rangle            &=\ds \frac{1}{\sqrt{2^n}} |x_0\rangle+\frac{1}{\sqrt{2^n}} \sum_{x\neq x_0}|x\rangle, \\
|b\rangle &= \ds \frac{1}{\sqrt{2^n}} \sum_{x=0}^{2^n-1}(-1)^{f(x)}|x\rangle &=\ds \frac{1}{\sqrt{2^n}} |x_0\rangle-\frac{1}{\sqrt{2^n}} \sum_{x\neq x_0}|x\rangle.
\ea\]
The state proportional to $|a\rangle+|b\rangle$ is $|x_0\rangle$, which can be obtained in time $O(n/\epsilon)$ by lemma \ref{lem-lcs}. The angle $\theta$ between $|a\rangle$ and $|b\rangle$ is about $\pi-1/\sqrt{2^n}$. Simple analysis about the error shows that, we should choose $\epsilon=\sqrt{2^n}$. This is the same result as Grover's algorithm.

\begin{figure}[H]
  \centering
  \includegraphics[width=6cm]{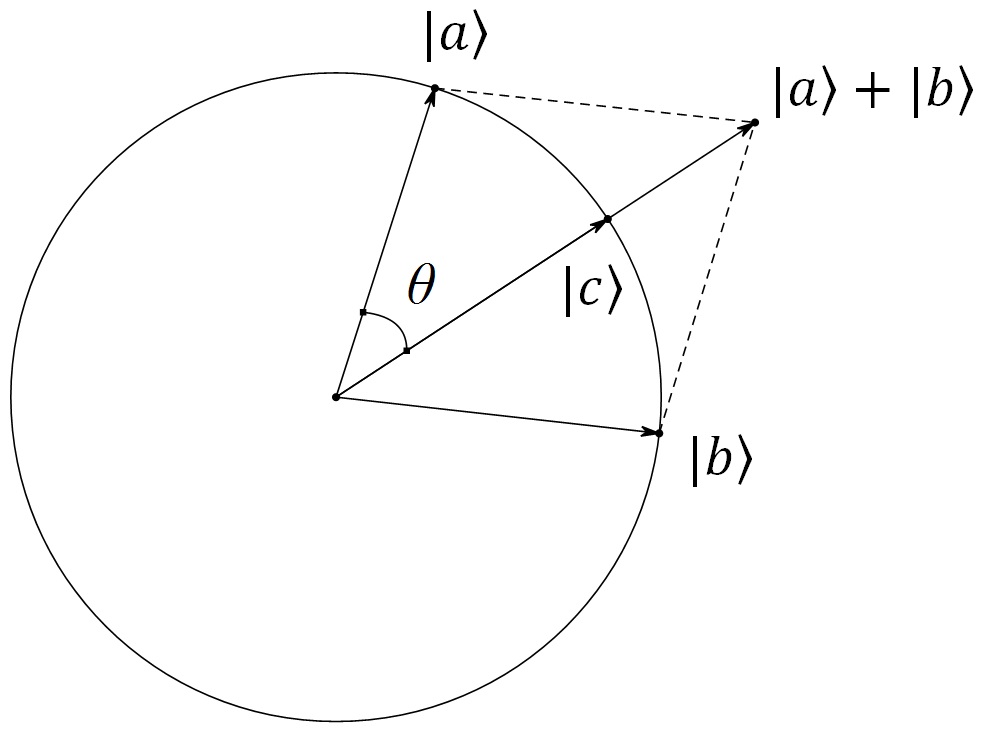}\\
  \caption{Linear combination of two quantum states}
  \label{lcs}
\end{figure}

With the above result in lemma \ref{lem-lcs}, now we can study the problem of linear combination of quantum states.
First, we consider a simple case: the construction of $|c\rangle$ proportional to $|a\rangle+|b\rangle$.
Denote $\theta$ as the angle between $|a\rangle$ and $|c\rangle$ (see \eqref{lcs}), the anti-clockwise rotation with angle $\theta$ in the plane spanned by $|a\rangle,|b\rangle$ as $R_\theta$. Note that the angle between $|a\rangle$ and $|b\rangle$ equals $2\theta$ and the rotation
\be \label{rotation}
R_{4\theta}=(I-2|b\rangle\langle b|)(I-2|a\rangle\langle a|)
\ee
can be implemented in time $O(T_{\textmd{in}})$, so $R_\theta=R_{4\theta}^{1/4}$ can be implemented in time $O(T_{\textmd{in}}/\epsilon)$ too by lemma \ref{lem-lcs}. Finally, $|c\rangle=R_\theta|a\rangle$ can be obtained in time $O(T_{\textmd{in}}/\epsilon)$.

We can actually compute the angle $\theta$ from
\be \label{angle}
\cos \theta = \frac{\langle a| (|a\rangle+|b\rangle)}{\||a\rangle+|b\rangle\|}
=\frac{1+\langle a|b\rangle}{\sqrt{2+2\langle a|b\rangle}}.
\ee
By swap test, this value and so $\theta$ can be estimated to accuracy $\epsilon$ in time $O(T_{\textmd{in}}/\epsilon)$. From $\theta$, we can construct the rotation $R_\theta$ easily. Therefore, $|c\rangle$ can be obtained in time $O(T_{\textmd{in}}/\epsilon)$ to accuracy $\epsilon$.

Next, we consider a little more general case: To obtain $|c\rangle$ proportional to $\alpha|a\rangle+\beta|b\rangle$. Similarly, we denote the angle between $|a\rangle$ and $|c\rangle$ as $\theta$, the angle between $|a\rangle$ and $|b\rangle$ as $\phi$. Note that at this time $\phi$ may not equals $2\theta$. From a similar formula as \eqref{angle}, we can estimate $\theta,\phi$ in time $O(T_{\textmd{in}}/\epsilon)$ to accuracy $\epsilon$. Now we can set $\theta=r\phi$. The rotation $R_\phi$ has the expression \eqref{rotation}. From the assumption, $R_\theta=R_\phi^r$ and so $|c\rangle$ can be obtained in time $O(T_{\textmd{in}}/\epsilon)$.

Finally, we can consider the general case, that is obtaining $|y\rangle$ proportional to $\sum_{j=0}^{l-1}\alpha_j|v_j\rangle$. For simplicity, we suppose $l=2^m$. In order to obtain $|y\rangle$, first we can calculate the quantum states proportional to $\alpha_{2i}|v_{2i}\rangle+\alpha_{2i+1}|v_{2i+1}\rangle$ for all $i=0,1,\ldots,l/2-1$ in time $O(T_{\textmd{in}}/\epsilon)$. The corresponding norm can be estimated at the same time. Now we can continue the above study about the linear summation of new obtained $2^{m-1}$ quantum states. This will takes $m$ steps. So the final complexity is
$
\sum_{i=0}^{m-1} {2^i T_{\textmd{in}}}/{\epsilon^{m-i}} \approx {T_\textmd{in}}/{\epsilon^m} = {T_\textmd{in}}/{\epsilon^{\log l}}
= T_\textmd{in}l^{\log 1/\epsilon}.
$
Note that after $m$ steps, the error is enlarged into $l\epsilon$. So we can choose $l\epsilon=\epsilon_0$, that is $1/\epsilon=l/\epsilon_0$. Combining the above analysis, the complexity to obtain $|y\rangle$ to accuracy $\epsilon_0$ is $O(T_\textmd{in}l^{\log l/\epsilon_0})$. This above result is almost polynomial in $l$. Therefore, we have

\bl\label{lem:lin-com-new}
Given $l$ complex numbers $\alpha_j=r_je^{\i\theta_j}$ and $l$ quantum states $|v_j\rangle$, where $j=0,1,\ldots,$ $l-1$, then we can obtain the quantum state proportional to $\sum_{j=0}^{l-1}\alpha_j|v_j\rangle$ in time
$
O(T_\emph{in}l^{\log l/\epsilon})
$
to accuracy $\epsilon$, where $T_{\emph{in}}$ is the maximal complexity to prepare $|v_j\rangle$ for $j=0,1,\ldots,l-1$.
\el

\bl\label{LCU:general}
Given $2l$ quantum states $|\phi_j^{\pm}\rangle=\alpha_j|0\rangle|u_j\rangle\pm \beta_j|0\rangle|v_j\rangle$,
which can prepared in time $O(T_{\emph{in}})$.
Then the quantum state proportional to $\sum_j\alpha_j|u_j\rangle$ can be obtained in time
$
O(T_{\emph{in}} l^{\log (l/\epsilon)}/\epsilon)
$
to precision $\epsilon$.
\el

\bo
First, we apply swap test to estimate all $\alpha_j$. Then we obtain $|u_j\rangle$, which is proportional to $|\phi_j^+\rangle+|\phi_j^-\rangle$ by LCU given in lemma \ref{lem:lin-com-new} in time $O(T_{{\rm in}}/\epsilon)$. Finally, we again apply the LCU given in lemma \ref{lem:lin-com-new} to prepare the quantum state proportional to $\sum_j\alpha_j|u_j\rangle$.
\qed\eo

Note that in the HHL algorithm to solve linear system $Ax=b$, before measurement, we have a state in the form $|0\rangle A^{-1}|b\rangle+|0\rangle^\bot$, which is obtained in time $O(\kappa(\log n)^2/\epsilon)$. Similarly to the above method, we can get the the solution $|x\rangle=A^{-1}|b\rangle/\|A^{-1}|b\rangle\|$ in time $O(\kappa(\log n)^2/\epsilon^2)$, which is linear in the condition number. This idea is much simpler to reduce the condition number into linear than \cite{ambainis,childs-linear-system}.

\section{Quantum stationary iteration}
\setcounter{equation}{0}
\label{Quantum stationary iteration}

Before the study of quantum Arnoldi and conjugate gradient iteration method, in this section, we consider the following simple iteration method:
\be\label{qim:eq1}
x\mapsto Ax+b
\ee
with a given matrix $A$ and a given vector $b$. Such iteration method is one of the basic iteration method to solve linear systems, such as Jacobi iteration, Gauss-Seidel iteration and SOR \cite{golub}, \cite{saad}, \cite{wilkinson}.
After $\eta$ iterations with initial vector $x^{(0)}$, we will obtain a target vector
\be\label{qim:eq2}
y=A^\eta x^{(0)}+A^{\eta-1} b+\cdots+A b+b.
\ee
The question is do we have better method to get the quantum state $|y\rangle$?

Note that the iteration method converges if and only if the spectral radius of $A$ is strictly less than 1. So we assume that the quantum state $|x^{(0)}\rangle,|b\rangle$ can prepared efficiently and $A$ is Hermitian with singular values lie between $[1/\kappa,1)$. We also assume that $\|x^{(0)}\|=\|b\|=1$ for simplicity. This problem was first considered in \cite{kerenidis}. Because of the expression (\ref{qim:eq2}) of $y$, more generalization like the quantum state of $p(A)b$, for some given polynomial $p$, can be obtained similarly, which forms the basic idea of Krylov method.

\subsection{Quantum iteration method: I}

Denote the initial vector as $x^{(0)}$, then the classical iteration method can be described as:
\be\label{qim:eq3}
x^{(0)}\mapsto x^{(1)}=Ax^{(0)}+b\mapsto\cdots\mapsto x^{(\eta)}=Ax^{(\eta-1)}+b.
\ee
In each step, we can check whether or not $x^{(k)}$ is already good enough. If $x^{(k)}$ is already satisfying the desired condition, then the iteration method stops, otherwise it continues with initial vector $x^{(k)}$. Due to the No-Cloning Theorem, in quantum computer, if we performing a checking on the state $|x^{(k)}\rangle$ and if it does not satisfy the desired condition, then we should restart the iteration method with initial vector $x^{(0)}$. Because of this, the complexity of iteration method in quantum computer is exponentially depending on the number of iteration step $\eta.$ In this iteration method, we actually do not need to consider the procedure (\ref{qim:eq3}) directly, however, in many more complicate iteration methods such as gradient descent method and Newton's method considered in \cite{rebentros}, $A$ is depending on the result $x^{(k)}$ in the $k$-th iteration. Therefore, in order to do further generalizations, it also worth to study the iteration method (\ref{qim:eq3}) clearly.

In formula (\ref{hhl:eq2}) with initial state $|0,x\rangle$ and $f(x)=x$, we can get the following state
\be\label{qim:eq4}
|0\rangle A|x^{(0)}\rangle+|1\rangle|g\rangle,
\ee
for some unwanted state $|g\rangle$. Then applying quantum state linear combination method (see lemma \ref{lem:lin-com}), we can get the following state
\be\ba{lll}\label{qim:eq5} \vspace{.2cm}
&& \ds\frac{1}{\sqrt{2}}\Big(|0\rangle\Big(|0\rangle A|x^{(0)}\rangle+|1\rangle|g\rangle\Big)+|1,0\rangle|b\rangle\Big) \\
&\mapsto& \ds\frac{1}{2}|0,0\rangle (A|x^{(0)}\rangle+|b\rangle)
+\frac{1}{2}|1,0\rangle (A|x^{(0)}\rangle-|b\rangle)+\frac{1}{\sqrt{2}}|0,1\rangle|g\rangle.
\ea\ee
Combining (\ref{qim:eq4}) and (\ref{qim:eq5}), we have a quantum procedure to achieve
\be\label{qim:eq6}
|x\rangle\mapsto \frac{1}{2}|0\rangle (A|x^{(0)}\rangle+|b\rangle)+|1\rangle|g_1\rangle,
\ee
with some garbage state $|g_1\rangle$. Generalize this, we have

\bl
For any given quantum state $\alpha|0\rangle|u\rangle+|1\rangle|v\rangle$ and $|w\rangle$, we have a quantum algorithm to achieve the quantum state
\be\label{qim:eq7}
\frac{\alpha}{\sqrt{2(1+\alpha^2)}}|0\rangle(|u\rangle+|w\rangle)+|1\rangle|v'\rangle.
\ee
\el
\bl
A sequence with the iteration relation $\alpha_n\mapsto\frac{\alpha_n}{\sqrt{2(1+\alpha_n^2)}}$ with initial value $\alpha_0=1$ has the formula
$\alpha_n=1/\sqrt{2^{n+1}+2^n-2}$.
\el
\emph{Proof.} Denote $\beta_n=\alpha_n^{-2}$, then we have the relation $\beta_{n+1}=2\beta_n+2=2^{n+1}\beta_0+2^{n+1}+\cdots+2=2^{n+2}+2^{n+1}-2$. And so
$\alpha_n=1/\sqrt{2^{n+1}+2^n-2}.$
\hfill $\square$ \vspace{.2cm}

After $\eta$ steps of iteration, we finally obtain the quantum state
\be\label{qim:eq8}
\frac{\|x^{(\eta)}\|}{\sqrt{2^{\eta+1}+2^\eta-2}}|0\rangle |x^{(\eta)}\rangle+|1\rangle|g_\eta\rangle.
\ee
The probability to obtain $|x^{(\eta)}\rangle$ is
\be\label{qim:eq9}
\frac{\|x^{(\eta)}\|^2}{2^{\eta+1}+2^\eta-2}.
\ee
Before estimate the complexity of obtaining $|x^{(\eta)}\rangle$, we should estimate the error in each step. Denote the obtained vector in the $k$-th step as $\tilde{x}^{(k)}$.
If $\|x^{(k)}-\tilde{x}^{(k)}\|\leq \delta$, then $\|Ax^{(k)}-A\tilde{x}^{(k)}\|\leq \|A\|\delta\leq \delta$. In the $(k+1)$-th step, we will obtain an approximate $\tilde{x}^{(k+1)}$ of $A\tilde{x}^{(k)}$ such that $\|\tilde{x}^{(k+1)}-A\tilde{x}^{(k)}\|\leq \epsilon$, then
$\|x^{(k+1)}-\tilde{x}^{(k+1)}\| \leq \|Ax^{(k)}-A\tilde{x}^{(k)}\|+\|A\tilde{x}^{(k)}-\tilde{x}^{(k+1)}\| \leq \delta+\epsilon.$
Therefore, the final error between $x^{(\eta)}$ and the quantum state $\tilde{x}^{(\eta)}$ obtained in (\ref{qim:eq8}) is bounded by $\eta\epsilon$.
After normalization, $\||x^{(\eta)}\rangle-|\tilde{x}^{(\eta)}\rangle\|\leq \eta\epsilon/\|x^{(\eta)}\|$. Hence, we should choose $\epsilon$ as $\epsilon\|x^{(\eta)}\|/\eta$. And the complexity is $\widetilde{O}(\eta^2\sqrt{2^\eta}/\|x^{(\eta)}\|^2\epsilon)$, here $\widetilde{O}(\eta/\epsilon)$ is the complexity of matrix multiplication from $|x^{(0)}\rangle$ to $|x^{(\eta)}\rangle$ (see the analysis in proposition \ref{thm1}).

\subsection{Quantum iteration method: II}

The exponential dependence on the number of iteration step $\eta$ is due to the linear combination procedure (\ref{qim:eq3}) of quantum states. Actually, we have a better method to overcome this problem. Since $x^{(1)}=Ax^{(0)}+b=[A,I]\left(
                                               \begin{array}{c}
                                                 x^{(0)} \\
                                                 b \\
                                               \end{array}
                                             \right)$, here $I$ is identity matrix of suitable dimension.
So from $x^{(0)}$ to $x^{(1)}$, we can apply the matrix multiplication of $[A,I]$ and $\left(
                                               \begin{array}{c}
                                                 x^{(0)} \\
                                                 b \\
                                               \end{array}
                                             \right)$.
Just the same way as (\ref{qim:eq4}), we have
\be\ba{lll} \vspace{.2cm}\label{qim:eq10}
|(x^{(0)},b)\rangle &\mapsto& |0\rangle (A|x^{(0)}\rangle+|b\rangle)+|1\rangle|g\rangle \\
&=& \|x^{(1)}\| |0\rangle |x^{(1)}\rangle+|1\rangle|g\rangle.
\ea\ee
Then we also need to obtain $|(x^{(1)},b)\rangle$ by adding $|b\rangle$ into the obtained quantum state in (\ref{qim:eq10}), which induces the same thing happened in the above method.
Instead, we can choose the initial state as
\be\label{qim:eq11}
|(x^{(0)},\underbrace{b,b,\ldots,b}_\eta)\rangle=: |(x^{(0)},b^\eta)\rangle.
\ee
And in each iteration step, we can choose use the matrix
$\left(
   \begin{array}{cc}
     [A,I] & 0 \\
     0 & I \\
   \end{array}
 \right)$. So we will have the following procedure
\be\label{qim:eq12}
|(x^{(0)},b^\eta)\rangle\mapsto |(x^{(1)},b^{\eta-1})\rangle\mapsto\cdots\mapsto|x^{(\eta)}\rangle.
\ee
The error estimation is the same as above, so the complexity will be $\widetilde{O}(\eta^2/\|x^{(\eta)}\|^2\epsilon)$.

\subsection{Quantum iteration method: III}

The method introduced above by extending the given matrix may appear some other problems, such that the extended matrix may not satisfy the assumptions as $A$ did.
This can be solved in the following way within the same complexity.
In formula (\ref{hhl:eq2}) with initial state $|0,x^{(0)}\rangle,|0,b\rangle$ respectively, we will have
\be\ba{lll} \vspace{.2cm}\label{qim:eq13}
|\psi_{\eta}\rangle &=& A^{\eta}|x^{(0)}\rangle|0\rangle+|G_\eta\rangle|1\rangle, \\
|\psi_k\rangle &=& A^k|b\rangle|0\rangle+|G_k\rangle|1\rangle,~(0\leq k\leq \eta-1),
\ea\ee
for some garbage states $|G_k\rangle~(0\leq k\leq \eta)$.
Similarly, we can get the summation of them:
\be\label{qim:eq14}
\frac{1}{\eta+1}|0\rangle \Big(A^{\eta}|x^{(0)}\rangle+\sum_{k=0}^{\eta-1}A^k|b\rangle\Big)+|\textmd{others}\rangle.
\ee
Note that the error between $A^{\eta}|x^{(0)}\rangle$ can the obtained quantum state is bounded by $\eta \epsilon$ because of (\ref{error}). The error between $A^k|b\rangle$ and the obtained quantum state is bounded by $k\epsilon$. Finally, the error between $x^{(\eta)}$ and the obtained quantum state is bounded by $\eta^2\epsilon$. The error between the normalized vectors is bounded by $\eta^2\epsilon/\|x^{(\eta)}\|$. Therefore, the complexity to get $|x^{(\eta)}\rangle$ is $\widetilde{O}(\eta^3/\|x^{(\eta)}\|^2\epsilon)$. The method considered in \cite{kerenidis} belongs to this category.

The following table is a conclusion of the complexity of the above three quantum iteration methods to the iteration method (\ref{qim:eq1}).

{\renewcommand\arraystretch{1.7}
\begin{table}[H]
\centering\caption{Comparison of the complexity of different quantum iteration methods}
\begin{tabular}{c|c}
  \hline
   \hspace{2cm}Method \hspace{2cm} & \hspace{2cm} Complexity  \hspace{2cm} \\ \hline
  I   & $\widetilde{O}(\eta^2\sqrt{2^\eta}/\|x^{(\eta)}\|^2\epsilon)$ \\
  II  & $\widetilde{O}(\eta^2/\|x^{(\eta)}\|^2\epsilon)$ \\
  III & $\widetilde{O}(\eta^3/\|x^{(\eta)}\|^2\epsilon)$ \\
  \hline
\end{tabular}
\end{table}}

\section{Quantum Arnoldi iteration}
\label{Quantum Arnoldi iteration}

In numerical linear algebra, the Arnoldi iteration \cite{arnoldi}, \cite{saad} is an important iterative method to approximate eigenvalues of large sparse matrices. It can also applied to approximate the solution of large sparse linear systems in a low dimension. In this section, we are denoting to extend it into a quantum algorithm with better efficiency.

Recall that, Hamiltonian simulation together with quantum phase estimation can be used to estimate the eigenvalues of Hermitian matrices \cite{abrams}, however, they cannot tell any information about the eigenvalues for non Hermitian matrices. So quantum Arnoldi iteration, if contains a high efficiency, will be a good quantum algorithm to estimate eigenvalues of non Hermitian matrices.

In the following, first, we will review the classical Arnoldi iteration method; then we will give two versions of quantum Arnoldi iteration method. The first one is a direct generalization, whose complexity is exponentially depends on the iteration steps. The modified one applies the property of Krylov and the new method of linear combination of quantum states, which is polynomial in the iteration steps.

\subsection{Classical Arnoldi iteration}
\label{Classical Arnoldi iteration}

Let $A$ be a $n\times n$ matrix, Arnoldi iteration is a mean to find a unitary matrix $X$ such that $X^\dagger AX$ is a Hessenberg matrix. The basic algorithm is as follows

\begin{algorithm}
\caption{Arnoldi iteration \cite{saad}}
\begin{algorithmic}[1]
\STATE Choose a vector $|x_0\rangle$
\STATE For $k=0,1,2,\ldots,m-1$ do
\STATE \hspace{.2cm} Compute $h_{ik}=\langle x_i|A|x_k\rangle$ for $i=0,1,\ldots,k$
\STATE \hspace{.2cm} Compute $w_{k+1}=A|x_k\rangle-\sum_{i=0}^k h_{ik}|x_i\rangle$
\STATE \hspace{.2cm} Compute $h_{k+1,k}=\|w_{j+1}\|_2$
\STATE \hspace{.2cm} If $h_{k+1,k}=0$ then stop, else $|x_{k+1}\rangle=w_{k+1}/h_{k+1,k}$
\STATE End do
\end{algorithmic}
\end{algorithm}

The vectors $\{|x_0\rangle,|x_1\rangle,\ldots,|x_{m-1}\rangle\}$ is an orthogonal basis of the Krylov subspace spanned by $|x_0\rangle, A|x_0\rangle, $ $\ldots,A^{m-1}|x_0\rangle$. Moreover, it is the result of Gram-Schmidt orthogonalization. Denote $X_m=[|x_0\rangle,|x_1\rangle,\ldots,|x_{m-1}\rangle]$ $=\sum_{k=0}^{m-1}|x_k\rangle\langle j|$ which is a $n\times m$ matrix whose columns are generated by $|x_k\rangle$. Let $\widetilde{H}=(h_{ij})_{(m+1)\times m}$ be the matrix generated by all $h_{ij}$. And set $H_m$ as the $m\times m$ matrix obtained by deleting the last row of $\widetilde{H}$. The last row of $\widetilde{H}$ is $h_{m,m-1}(0,0,\ldots,0,1)=h_{m,m-1}\langle m|$. Then in algorithm 1 after $m$ steps, we have
\[
A|x_k\rangle=\sum_{i=0}^{k+1} h_{ik}|x_i\rangle.
\]
That is
\be
AX_m=\left(
              \begin{array}{cc}
                X_m & |x_m\rangle \\
               \end{array}
              \right) \left(
                            \begin{array}{c}
                             H \\
                             h_{m,m-1}\langle m| \\
                            \end{array}
                        \right)=X_mH+h_{m,m-1}|x_m\rangle\langle m|.
\ee
So $X_m^\dagger A X_m=H_m$.

\br
(1). If $\lambda$ is an eigenvalue of $H$ with eigenvector $|u\rangle$, then
\be
AX_m|u\rangle=X_mH|u\rangle+h_{m,m-1}\langle m|u\rangle|x_m\rangle=\lambda X_m|u\rangle+h_{m,m-1}\langle m|u\rangle|x_m\rangle.
\ee
When $h_{m,m-1}$ is small, then $\lambda$ can be viewed as an approximate of the eigenvalue of $A$ with eigenvector $X_m|u\rangle$. So $(\lambda,X_m|u\rangle)$ is a Ritz pair of $A$ (see \cite{golub}).

(2). Consider the linear system $Ax=b$. Assume that $x=X_my$, then we will get $Hy=X_m^\dagger A X_my=X^\dagger b$. So we can first solve the linear system $Hy=X_m^\dagger b$, then recover the original solution from $x=X_my$. Note that, generally  $x=X_my$ may not hold exactly, however, we can expect there is a $y$ such that $X_my$ is a good approximate of $x$ as $m$ grows. This means, we want to find an approximate solution of $Ax=b$ in a low dimensional subspace. When considering solve the linear system, the initial vector is often choose as proportional to $|b\rangle-A|x_0'\rangle$ for some guess $|x_0'\rangle$ of the solution.
\er

About $X_k$, we can only get the quantum state of $|x_k\rangle$. Reading out will takes at leats $O(n)$ steps. However, as for $H$,  all the entries are given in the inner product form, so we can get the classical information of $H$ directly by swap test. Assume that we already get $|x_0\rangle,\ldots,|x_{m-1}\rangle$ in time $O(C_0),\ldots,O(C_{m-1})$ respectively, and get $H$ in time $O(C_H)$. Generally $m$ is small, so the eigenvalues of $H$ can be computed efficiently. As for solving the linear system $Hy=X^\dagger b$. The quantum state of $X^\dagger b$ is proportional to $X_m^\dagger |b\rangle=\sum_{k=0}^{m-1}\langle b|x_j\rangle|j\rangle$. As shown in corollary \ref{cor:inner product}, $\langle b|x_j\rangle$ can be computed in time $O(C_j/\epsilon)$. And so the quantum state $X_m^\dagger |b\rangle$ can be obtained efficiently when $m$ is small. Therefore, the solution $y$ and so $|y\rangle$ of $Hy=X^\dagger b$ can be obtained in time
\be\label{analysis1}
O\left(C_H+\sum_{j=0}^{m-1} C_j/\epsilon\right).
\ee
Since $x=X_my$, we have $|x\rangle=\sum_j\langle j|y\rangle |x_j\rangle=\sum_j y_j |x_j\rangle$. Lemma \ref{lem:lin-com} shows that this can be obtained in time
\be\ba{lll}\label{analysis2}
\ds O\left(\max_{0\leq j\leq m-1} C_j m\max_{0\leq j\leq m-1} |y_j|\left\|\sum_{j=0}^{m-1} y_j |x_j\rangle\right\|_2^{-1}\right)
&=& \ds O\left(\max_{0\leq j\leq m-1} C_j m\||y\rangle\|_2^{-1}\right) \\
&=& \ds O\left(\max_{0\leq j\leq m-1} C_j m\right),
\ea\ee
due to the orthogonality of $|x_j\rangle$. Therefore, in the following, we just need to focus on the estimation of $C_j$ and $C_H$.

\subsection{Quantum Arnoldi iteration: direct generalization}

Since in the numerical case, we cannot expect that $h_{k+1,k}=0$ exactly. We give a bound $\delta$ about the norm and assume that the algorithm stops when $h_{k+1,k}\leq \delta$. Assume that we already have $|x_0\rangle,\ldots,|x_k\rangle$ which obtained in time $O(C_0),\ldots,O(C_k)$ respectively. Then in order to estimate $h_{ik}=\langle x_i|A|x_k\rangle$, we assume that the singular values of $A$ lie between $1/\kappa$ and 1, the Hamiltonian simulation of
$\left(
\begin{array}{cc}
0 & A \\
A^\dagger & 0 \\
\end{array}
\right)$ can be implemented efficiently.
Then $|0\rangle A|x_k\rangle+|0\rangle^\bot$ can be prepared in time $O(C_k(\log n)^2/\epsilon)$ by proposition \ref{thm1}. By corollary \ref{cor:inner product}, the inner product of $|0\rangle|x_i\rangle$ and $|0\rangle A|x_k\rangle$, which equals $h_{ik}$, can be estimated in time
\be \label{entries of H:complexity}
O(\max\{C_i,C_k(\log n)^2/\epsilon\}/\epsilon)=O(C_k(\log n)^2/\epsilon^2),
\ee
since the preparation of $|x_k\rangle$ relies on $|x_i\rangle$ for $k\geq i$, which implies $C_k\geq C_i$. Whence we obtain $h_{ik}$, it belongs to classical information and so can be used as many times as we want.

Assume that the singular value decomposition of $A=\sum_l \sigma_l|u_l\rangle\langle v_l|$. Set $|x_j\rangle=\sum_l x_{jl}|v_l\rangle$, then
$|h_{ij}|\leq |A|x_j\rangle|\leq 1$. Before the algorithm stops, we always have $h_{k+1,k}\geq \delta$.
To construct $|x_{k+1}\rangle$, by lemma \ref{lem:lin-com} with $l=k+1$, $|v_0\rangle=|0\rangle A|x_k\rangle+|0\rangle^\bot$ and $|v_i\rangle=|0\rangle|x_i\rangle$. The coefficient $\alpha_0=1$ and $\alpha_i=-h_{ik}$. Then we can obtain $|x_{k+1}\rangle$ in time
\be
O(C_{k+1})=O(C_k(\log n)^2 (k+1)/\epsilon \delta).
\ee
By induction and $C_0=1$,
\be \label{complexity of x(k+1)}
O(C_{k+1}) = O((k+1)!(\log n)^{2(k+1)}/\delta^{k+1}\epsilon^{k+1}).
\ee

\bp \label{result1}
The quantum state $|x_k\rangle$ can be prepared in time $O(k!(\log n)^{2k}/\delta^{k}\epsilon^k)$.
\ep

\bp \label{result2}
The matrix $H$ can be obtained in time $O((m+1)!(\log n)^{2m}/\delta^{m-1}\epsilon^{m+1})$.
\ep

\bo
By \eqref{entries of H:complexity} and \eqref{complexity of x(k+1)}, the entry $h_{ik}$ of $H$ can be obtained in time
$O(C_k(\log n)^2/\epsilon^2)=O(k!(\log n)^{2(k+1)}/\delta^{k}\epsilon^{k+2}).$ For any fixed $k$, each $i$ ranges from 0 to $k$. Totally, the complexity to obtained $H$ is
\begin{displaymath}
O\left(\sum_{k=0}^{m-1} (k+1)!(\log n)^{2(k+1)}/\delta^{k}\epsilon^{k+2}\right)=O((m+1)!(\log n)^{2m}/\delta^{m-1}\epsilon^{m+1}).
\end{displaymath}
\eo

Now we assume that $m$ is a small constant. As analyzed in the final part of subsection \ref{Classical Arnoldi iteration}, we have

\bp \label{result3}
Let $A$ be a $n\times n$ matrix. Assume that the singular values of $A$ lie between $1/\kappa$ and 1, the Hamiltonian simulation of
$\left(
\begin{array}{cc}
0 & A \\
A^\dagger & 0 \\
\end{array}
\right)$ can be implemented efficiently. Then

(1). $m$ eigenvalues and eigenvectors of $A$ can be computed in time
$
O((\log n)^{2m}/\delta^{m-1}\epsilon^{m+1}).
$

(2). The quantum state of the solution of $Ax=b$ can be obtained in time
$
O((\log n)^{2(m-1)}/\delta^{m-1}\epsilon^{m-1}).
$
\ep

\bo
By applying any classical algorithm (like QR algorithm) to $H$, we can compute all the eigenvalues of $H$ in time $O(m^3)$. So the $m$ approximate eigenvalues of $A$ can be obtained in time $O(m^3+(m+1)!(\log n)^{2m}/\delta^{m-1}\epsilon^{m+1})=O((\log n)^{2m}/\delta^{m-1}\epsilon^{m+1})$ if we assume that $m$ is a constant. From the analysis in \eqref{analysis2}, the solution of $Ax=b$ can be obtained in  $O(m!(\log n)^{2(m-1)}/\delta^{m-1}\epsilon^{m-1})$ $=O((\log n)^{2(m-1)}/\delta^{m-1}\epsilon^{m-1})$ when $m$ is a constant.
\eo

The quantum Arnoldi method obtained by direct generalization is not so good, the complexity not only exponentially depends on the iteration step, but also contains a factorial term about the iteration step. In the next subsection, we will improve this algorithm.

\subsection{Quantum Arnoldi iteration: improved algorithm}

The exponential dependence on $m$ in proposition \ref{result1}, \ref{result2}, \ref{result3} comes from the preparation of $|x_{k}\rangle$ which instead needs many copies of $|x_0\rangle,\ldots,|x_{k-1}\rangle$ and exponentially many copies of $|x_0\rangle$. Since $|x_{k}\rangle=q_{k}(A)|x_{0}\rangle$ for some polynomial $q_{k}$, if we can construct this polynomial in each step, then we can prepare $|x_{k}\rangle$ more efficiently. In this subsection, we will show an improved quantum version of Arnoldi iteration.

As for the Arnoldi iteration, we can set $|x_k\rangle=T_k|x_0\rangle$, where
\be \label{tj:def}
T_k = \sum_{l=0}^{k} \alpha_{kl} A^l,
\ee
for some parameter $\alpha_{kl}$ depends on $H$.

Since
\[
w_{k+1} = \ds A|x_k\rangle - \sum_{i=0}^k h_{ik} |x_i\rangle
= \ds AT_k |x_0\rangle - \sum_{i=0}^k h_{ik} T_i |x_0\rangle
= \ds \Bigg( AT_k  - \sum_{i=0}^k h_{ik} T_i \Bigg) |x_0\rangle
\]
and $h_{k+1,k}=\|w_{k+1}\|_2$, we have
\be \label{tj:recursive-formula}
T_{k+1} = \frac{1}{h_{k+1,k}}  \Bigg( AT_k  - \sum_{i=0}^k h_{ik} T_i \Bigg).
\ee

Note that $T_0$ is identity. By formula \eqref{tj:def} and \eqref{tj:recursive-formula}, we have
\[\ba{lll} \vspace{.2cm}
T_{k+1} &=& \ds \frac{1}{h_{k+1,k}} \Bigg(\sum_{l=0}^{k} \alpha_{kl}A^{l+1} - \sum_{i=0}^k h_{ik} \sum_{l=0}^{i} \alpha_{il} A^l \Bigg) \\ \vspace{.2cm}
&=&  \ds \frac{1}{h_{k+1,k}} \Bigg(\sum_{l=1}^{k+1} \alpha_{k,l-1}A^{l} - \sum_{l=0}^{k} \Big(\sum_{i=l}^k h_{ik} \alpha_{il}\Big) A^l \Bigg) \\
&=&  \ds \frac{1}{h_{k+1,k}} \Bigg(
-\Big(\sum_{i=0}^k h_{ik} \alpha_{i0}\Big)
+ \sum_{l=1}^{k} \Big(\alpha_{k,l-1}-\sum_{i=l}^k h_{ik} \alpha_{il}\Big)A^l
+ \alpha_{k,k}A^{k+1}   \Bigg).
\ea\]
Finally, we get
\be  \label{tj:coef}
\alpha_{k+1,l} =  \frac{1}{h_{k+1,k}} \Big(\alpha_{k,l-1}-\sum_{i=l}^k h_{ik} \alpha_{il}\Big),
\hspace{.5cm} (0\leq l \leq k+1).
\ee

For simplicity, we set $\beta_{k+1,l} = \alpha_{k+1,l} h_{k+1,k}=\alpha_{k,l-1}-\sum_{i=l}^k h_{ik} \alpha_{il}$. Until now, we have obtain a recursive formula \eqref{tj:coef} about the coefficients of the polynomial $q_k(A)$.

Assume that we already have $T_0,\ldots,T_k$, which means we already have $\alpha_{ij},h_{{ij}}$ for $0\leq i,j\leq k$, which are classical data.  Note that it may take a lot to calculate $\alpha_{ij},h_{{ij}}$, however, whence we obtain them, the complexity to compute $\beta_{k+1,l}$ is just $O(k+1)$. Now set
$
S_{k+1} = \sum_{l=0}^{k+1} \beta_{k+1,l} A^l.
$
Then $S_{k+1}$ can be obtained in time $O((k+1)(k+2))$, since there are $k+2$ coefficients we should compute. Also $|x_{k+1}\rangle$ is proportional to $S_{k+1}|x_0\rangle$. The quantum state $|0\rangle A^l |x_0\rangle +|0\rangle^\bot$ can be obtained in time $O(l(\log n)^2/\epsilon)$ due to proposition \ref{thm1}. By lemma \ref{lem:lin-com}, the quantum state of $|x_{k+1}\rangle$ can be obtained in time
\be \label{improved-complexity1}
O( (k+1)(k+2)+ (k+1)(\log n)^2 \max_{0\leq l \leq k+1} |\beta_{k+1,l}|  /\epsilon \delta).
\ee
By induction on \eqref{tj:coef} and note that $h_{ij}\leq 1$, we have $|\beta_{k+1,l}| \leq (k+1)!/\delta^{k+1}$. Therefore, \eqref{improved-complexity1} can be changed into
\be \label{improved-complexity2}
O( (k+2)! (\log n)^2/\epsilon \delta^{k+2}).
\ee
Since the preparation of $|x_{k+1}\rangle$ depends on $\alpha_{ij},h_{{ij}}$, which further depend on the preparation of $|x_0\rangle,\ldots,|x_k\rangle$, which means the actually required time to prepare $|x_{k+1}\rangle$ is
\be \label{improved-complexity2:true}
O\left(\sum_{l=0}^{k+1} (l+1)! (\log n)^2/\epsilon \delta^{l+1}\right)
=O( (k+3)! (\log n)^2/\epsilon \delta^{k+2}  ).
\ee

Then the complexity to calculate $h_{k+1,l}$ is
$
O( (k+3)! (\log n)^2/\epsilon^2 \delta^{k+2}).
$
So the matrix $H$ can be obtained in time
\be \label{improved-complexity4}
O\Bigg (\sum_{k=0}^{m-1}  (k+1)(k+3)! (\log n)^2/\epsilon^2 \delta^{k+2} \Bigg)
=O((m+5)!(\log n)^2/\epsilon^2 \delta^{m+1}).
\ee
Summarize the above analysis, we have
\bp \label{new:result}
The quantum state $|x_k\rangle$ for all $0\leq k\leq m-1$ can be obtained in time $O( (k+3)! (\log n)^2/\epsilon^2 \delta^{k+2})$ and the matrix $H$ can be obtained in time $O((m+5)!(\log n)^2/\epsilon^2 \delta^{m+1}).$ Moreover, the linear system $Ax=b$ can be solved in time  $O((m+6)!(\log n)^2/\epsilon^2 \delta^{m+1}).$
\ep

Compared with proposition \ref{result1} and \ref{result2}, the above results are much better, although it contains a factor $(m+6)!$ and $1/ \delta^{m+1}$. These two factors come from the estimation of $\max_{0\leq l \leq k+1} |\beta_{k+1,l}|$.
Note that in formula \eqref{improved-complexity1} and \eqref{improved-complexity2}, we replace $\max_{0\leq l \leq k+1} |\beta_{k+1,l}|$ by $(k+1)!/\delta^{k+1}$. However, in the way to prepare $|x_{k+1}\rangle$, we should compute all the values of $\beta_{k+1,l}$, so the maximum of $|\beta_{k+1,l}|$ may not achieve $(k+1)!$ in practice. Considering about this influence, the linear combination method about quantum states proposed in lemma \ref{LCU:general} will play an important role now, since it is independent of the coefficients.

By lemma \ref{LCU:general} and similar to the analysis of \eqref{improved-complexity2:true}, the quantum state $|x_{k+1}\rangle$ can actually obtained in time
$O(k^{1+\log (k/\epsilon)}(\log n)^2/\epsilon^2)$. Since $h_{ik}=\langle x_i|A|x_k\rangle$, by applying swap test on $|0\rangle|x_i\rangle$ and $|0\rangle A|x_k\rangle+|0\rangle^\bot$ (see proposition \ref{thm1}), we can estimate $h_{ik}$ in time $O(k^{1+\log (k/\epsilon)}(\log n)^2 /\epsilon^4)$ to precision $\epsilon$, here one $\epsilon$ comes from proposition \ref{thm1} and another one comes from swap test.
Therefore, all the entries (about $O(m^2)$ entries) of the matrix $H$ can be computed in time $O(m^{3+\log (m/\epsilon)}(\log n)^2 /\epsilon^4)$. Conclude this, we have

\bt \label{new:result1}
The quantum state $|x_k\rangle$ can be prepared in time $O(k^{1+\log (k/\epsilon)}(\log n)^2 /\epsilon^2)$ and
the matrix $H$ can be obtained in time $O(m^{3+\log (m/\epsilon)}(\log n)^2 /\epsilon^4)$. Moreover,
some extreme eigenvalues of $A$ and the solution of the linear system $Ax=b$ can be obtained in time $O(m^{3+\log (m/\epsilon)}(\log n)^2 /\epsilon^4)$.
\et

\bo Some extreme eigenvalues of $A$ can be obtained from the eigenvalues of $H$. By any classical algorithm, such as QR algorithm which costs $O(m^3)$, to compute the the eigenvalues of $H$, we will see that extreme eigenvalues of $A$ can be obtained in the same time as computing $H$.
From the analysis in \eqref{analysis1} and \eqref{analysis2}, we see that the solution of $Ax=b$ can be solved in time
$O(m C_{m-1}/\epsilon+C_H)=O(C_H)$.
\qed \eo

\section{Quantum conjugate gradient algorithm}
\label{Quantum conjugate gradient algorithm}

In mathematics, the conjugate gradient (CG) algorithm \cite{golub}, \cite{saad}, \cite{wilkinson} is one of the best known iterative techniques for solving symmetric positive-definite large sparse linear system $Ax=b$. It is a simplified and elegant variant of the symmetric Arnoldi method, i.e., Lanczos method. Just like Arnoldi method, conjugate gradient algorithm can also used to estimate the eigenvalues information of $A$, such as the largest and smallest eigenvalue of $A$.

\begin{algorithm}[H]
\caption{Conjugate gradient algorithm \cite{saad}}
\label{cg-alg}
\begin{algorithmic}[1]
\STATE Choose a vector $x_0$, set $r_0=b-Ax_0$ and $p_0=r_0$
\STATE For $k=0,1,2,\ldots$ do
\STATE \hspace{.2cm} If $r_k^Tr_k=0$ or $p_k^TAp_k=0$ then stops, else compute
\STATE \hspace{.4cm} $\alpha_k=r_k^Tr_k/p_k^TAp_k$
\STATE \hspace{.4cm} $x_{k+1}=x_k+\alpha_kp_k$
\STATE \hspace{.4cm} $r_{k+1}=r_k-\alpha_k Ap_k$
\STATE \hspace{.4cm} $\beta_k=r_{k+1}^Tr_{k+1}/r_k^Tr_k$
\STATE \hspace{.4cm} $p_{k+1}=r_{k+1}+\beta_k p_k$
\STATE End do
\end{algorithmic}
\end{algorithm}

As we can guess, if we generalize the above algorithm into a quantum algorithm directly, the complexity will be exponentially depends on the iteration step. So we intend to use a similar idea of quantum Arnoldi algorithm to amend this disadvantage.

For simplicity, we just choose $x_0=0$ and $r_0=p_0=|b\rangle$ are unit vectors. Also we set
\be \label{cg:eq1}
r_k = \ds \sum_{l=0}^k r_{kl} A^l |b\rangle, \hspace{.2cm}
x_k = \ds \sum_{l=0}^k x_{kl} A^l |b\rangle, \hspace{.2cm}
p_k = \ds \sum_{l=0}^k p_{kl} A^l |b\rangle.
\ee
We further assume that $r_{k,-1}=x_{k,-1}=p_{k,-1}=r_{k,k+1}=x_{k,k+1}=p_{k,k+1}=0$.
Then from the definition of CG algorithm (line 5, 6, 8), we have
\be\ba{lll} \vspace{.2cm} \label{cg:eq2}
x_{k+1} &=& x_k+\alpha_kp_k = \ds\sum_{l=0}^{k+1} (x_{kl}+\alpha_kp_{kl}) A^l|b\rangle, \\\vspace{.2cm}
r_{k+1} &=& r_k-\alpha_k Ap_k = \ds \sum_{l=0}^{k+1} (r_{kl}-\alpha_k p_{k,l-1})A^l|b\rangle, \\
p_{k+1} &=& r_{k+1}+\beta_k p_k = \ds \sum_{l=0}^{k+1} (r_{kl}-\alpha_k p_{k,l-1}+\beta_k p_{kl})A^l|b\rangle.
\ea\ee
This means
\be \label{cg:eq3}
x_{k+1,l} = x_{kl}+\alpha_kp_{kl}, \hspace{.2cm}
r_{k+1,l} = r_{kl}-\alpha_k p_{k,l-1}, \hspace{.2cm}
p_{k+1,l} = r_{kl}-\alpha_k p_{k,l-1}+\beta_k p_{kl}.
\ee

The initial values are $x_{00}=0,r_{00}=p_{00}=1$. Denote $\max_l|x_{kl}|=X_k,\max_l|r_{kl}|=R_k$ and $\max_l|p_{kl}|=P_k$. By lemma \ref{lem:lin-com}, the quantum states $|x_k\rangle,|r_k\rangle,|p_k\rangle$ can be prepared in time
\be \label{cg:eq4}
O((\log n)^2 k X_k/\|x_k\|\epsilon),~~~ O((\log n)^2 k R_k/\delta\epsilon),~~~ O((\log n)^2k P_k /\delta\epsilon)
\ee
respectively.

Before CG algorithm stops, $\alpha_k=r_k^Tr_k/p_k^T A p_k\leq 1/\delta^3$, since as residue $r_k^Tr_k$ decreases when $k$ grows. By \eqref{cg:eq3}, we have $|x_{k+1,l}|\leq |x_{kl}|+\alpha_k |p_{kl}|$, so
\be \label{cg:eq5}
X_{k+1}=X_k+P_k/\delta^3=\delta^{-3}\sum_{l=0}^k P_l.
\ee
Also from $|r_{k+1,l}|\leq |r_{kl}|+\alpha_k |p_{k,l-1}|$, we have
\be \label{cg:eq6}
R_{k+1}=R_k+P_k/\delta^3.
\ee
From $|p_{k+1,l}|\leq |r_{kl}|+\alpha_k |p_{k,l-1}|+\beta_k |p_{kl}|$, we have
\be \label{cg:eq7}
P_{k+1}=R_{k+1}+P_k=\sum_{l=0}^{k+1} R_l.
\ee

The initial values are $X_0=0,R_0=P_0=1$. From \eqref{cg:eq6} and \eqref{cg:eq7}, we have
$R_{k+1}=R_k+\delta^{-3}\sum_{l=0}^{k} R_l.$ Then $R_{k+1}=(2+\delta^{-3})R_k-R_{k-1}$. So
$
R_k=\lambda_1z_1^k+\lambda_2z_2^k,
$
where
\be\ba{lll}\vspace{.2cm} \label{cg:eq81}
&z_1=\ds\frac{2+\delta^{-3}+\sqrt{4\delta^{-3}+\delta^{-6}}}{2},& \hspace{1cm}
\lambda_1=\ds\frac{1+\delta^{-3}-z_2}{z_1-z_2}, \\
&z_2=\ds\frac{2+\delta^{-3}-\sqrt{4\delta^{-3}+\delta^{-6}}}{2},& \hspace{1cm}
\lambda_2=\ds\frac{1+\delta^{-3}-z_1}{z_2-z_1}.
\ea\ee
Hence
\be \label{cg:eq9}
P_k=\sum_{l=0}^k (\lambda_1z_1^l+\lambda_2z_2^l)
=\lambda_1 \frac{z_1^{k+1}-1}{z_1-1}+\lambda_2 \frac{z_2^{k+1}-1}{z_2-1},
\ee
and
\be\ba{lll}\vspace{.2cm} \label{cg:eq10}
X_k = \ds\delta^{-3}\sum_{l=0}^{k-1} P_l
&=& \ds\delta^{-3}\sum_{l=0}^{k-1} \Big(\lambda_1 \frac{z_1^{l+1}-1}{z_1-1}+\lambda_2 \frac{z_2^{l+1}-1}{z_2-1}\Big) \\
&=& \ds\frac{\delta^{-3}\lambda_1}{z_1-1}\Big(\frac{z_1^{k+1}-1}{z_1-1}-k-1\Big)
+\frac{\delta^{-3}\lambda_2}{z_2-1}\Big(\frac{z_2^{k+1}-1}{z_2-1}-k-1\Big).
\ea\ee

Note that the upper bounds $X_k,R_k,P_k$ may not achieve in specific examples. The above analysis are just in theory, and they show the worst cases.
From formula \eqref{cg:eq81}, it is easy to see that
$
z_1=O(\delta^{-3}),  z_2,  \lambda_1,   \lambda_2=O(1).
$
So
$
R_k, P_k, X_k=O(\delta^{-3k}).
$

\bp\label{cg:prop1}
The quantum state $|x_k\rangle$ of conjugate gradient algorithm and so the solution of the linear system $Ax=b$ can be obtained in time
$
O((\log n)^2 m^2 /\|x_m\|\epsilon\delta^{3m}),
$
where $m$ is the iteration steps of CG method.
\ep

The appearance of $m^2$ in the complexity is the same reason as \eqref{improved-complexity2:true}, that is the preparation of $|x_k\rangle$ depends on $|x_0\rangle,\ldots,|x_{k-1}\rangle$, so the complexity obtained in \eqref{cg:eq4} should added up as the complexity to prepare $|x_k\rangle$.
Just like HHL algorithm, we can substitute $1/\|x_m\|$ by $\kappa$, the condition number of $A$, so the complexity becomes $O( \kappa(\log n)^2 m^2 /\epsilon\delta^{3m}).$

The conjugate gradient algorithm can also used to approximate the eigenvalues of $A$ by considering the following traditional matrix \cite{saad}, whose eigenvalues can be computed efficiently by any classical eigenvalue algorithm when $m$ is small. The Hessenberg matrix obtained in Arnoldi method now reduces to the following tridiagonal matrix:
\be
T_m=\left(
      \begin{array}{ccccc}\vspace{.2cm}
        1/\alpha_0              &~~ \sqrt{\beta_0}/\alpha_0     &~~  &~~  &~~  \\\vspace{.2cm}
        \sqrt{\beta_0}/\alpha_0 &~~ 1/\alpha_1+\beta_0/\alpha_0 &~~ \sqrt{\beta_1}/\alpha_1 &~~  &~~  \\\vspace{.2cm}
         &~~ \ddots &~~ \ddots &~~\ddots  &~~  \\\vspace{.2cm}
         &~~  &~~ &~~ 1/\alpha_{m-2}+\beta_{m-3}/\alpha_{m-3} &~~  \sqrt{\beta_{m-2}}/\alpha_{m-2}  \\
         &~~  &~~ &~~ \sqrt{\beta_{m-2}}/\alpha_{m-2}  &~~  1/\alpha_{m-1}+\beta_{m-2}/\alpha_{m-2} \\
      \end{array}
    \right).
\ee
Each $\alpha_l,\beta_l$ can be computed in time $O((\log n)^2 l/\epsilon\delta^{3l+1})$. Therefore, the matrix $T_m$ can be constructed in time
\[
O\left(\sum_{l=0}^{m-1}(\log n)^2 l^2/\epsilon\delta^{3l+1}\right)
=O( (\log n)^2m^3/\epsilon\delta^{3m-2} ).
\]

\bp\label{cg:prop2}
In quantum computer, the largest and smallest eigenvalue of $A$ and so the condition number of $A$ can be computed in time
$
O( (\log n)^2m^3/\epsilon\delta^{3m-2} ).
$
\ep

Note that by lemma \ref{lem:lin-com-new}, the quantum states $|x_k\rangle,|r_k\rangle,|p_k\rangle$ can  prepared in time
$O(k^{1+\log k/\epsilon} (\log n)^2 /\epsilon \|x_k\|),$
$ O(k^{1+\log k/\epsilon} (\log n)^2 /\epsilon \delta)$ and
$O(k^{1+\log k/\epsilon} (\log n)^2 /\epsilon \delta)$ respectively. So,

\bt\label{cg:prop3}
Assume that CG algorithm stops at $m$ steps, then the solution of $Ax=b$ can be computed in time
$O(m^{1+\log m/\epsilon} (\log n)^2 /\epsilon \|x_m\|)$. Moreover,
the largest and smallest eigenvalue of $A$ can be obtained in time
$O( m^{2+\log m/\epsilon}(\log n)^2 /\epsilon^2 \delta)$.
\et

\section{Conclusions}
\label{conclusion}

In this paper, we have proposed the quantum versions of Arnoldi and CG iteration method. Under certain assumptions about quantum linear algebraic technique, these two methods contain a high performance than the classical methods. One important technique used in this paper is the linear combination of quantum states. It reduces the dependence of on iteration steps into almost polynomials. However, in this work, we did not concerns too much about the numerical stability. Since the classical numerical stable Arnoldi iteration, such as based on modified Gram-Schmidt or the Householder transformation are not easy to find their pretty good quantum versions. As an extension of this work, many Krylov methods such as the Lanczos method, preconditioned CG method can be studied similarly.

\end{document}